\documentclass[11pt]{iopart}

\usepackage{graphicx}
\usepackage{dcolumn}
\usepackage{bm}
\usepackage{algorithm}
\usepackage{enumerate}
\usepackage{algpseudocode}
\usepackage{algorithmicx}
\usepackage{acronym}
\usepackage{xcolor}
\usepackage[normalem]{ulem} 
\usepackage{amsthm, amssymb, amsfonts}
\usepackage{array}
\usepackage{caption}
\usepackage{subcaption}
\usepackage{hyperref}
\usepackage{cleveref}

\newcommand{\elliptica}{\texttt{Elliptica}}
\newcommand{\compose}{\texttt{CompOSE}}

\newcommand{\Msun}{\mathrm{\, M_\odot}}

\newcommand{\step}[1]{\textit{Step #1}}
\newcommand{\eq}[1]{Eq.~(\ref{#1})}
\newcommand{\fig}[1]{Fig.~(\ref{#1})}

\newcommand{\added}[1]{#1}
\newcommand{\rmed}[1]{}
\newcommand{\addedp}[1]{#1}

\newacro{ID}{Initial Data}
\newacro{BNS}{Binary Neutron Star}
\newacro{NS}{Neutron Star}
\newacro{BH}{Black Hole}
\newacro{EOS}{Equation of State}
\newacro{NR}{Numerical Relativity}
\newacro{PDE}{Partial Differential Equation}
\newacro{GW}{Gravitational Wave}
\newacro{BC}{Boundary Condition}
\newacro{XCTS}{Extended Conformal Thin Sandwich}
\newacro{TOV}{Tolman-Oppenheimer-Volko}
\acrodefplural{BC}{Boundary Conditions}
\acrodefplural{PDE}{Partial Differential Equations}
\acrodefplural{GW}{Gravitational Waves}
\acrodefplural{EOS}{Equation of States}
\acrodefplural{NS}{Neutron Stars}

\begin{document}

\title{Realistic binary neutron star initial data with \elliptica}

\author{Alireza Rashti$^{1,2}$\footnote{Author to whom any correspondence should be addressed.},
Andrew Noe$^{1,2}$}
\address{%
${}^{1}$ Institute for Gravitation \& the Cosmos, The Pennsylvania State
University, University Park PA 16802, USA \\
${}^{2}$ Department of Physics, The Pennsylvania State University, University
Park, PA 16802, USA
}%

\ead{numerical.relativity@gmail.com}

\begin{abstract}

This work introduces the \elliptica{} pseudo-spectral code for generating initial data 
of binary neutron star systems. Building upon the recent \elliptica{} code update, 
we can now construct initial data using not only piecewise polytropic equations of 
state, but also tabulated equations of state for these binary systems.
Furthermore, the code allows us to endow neutron stars within the binary system 
with spins. These spins can have a magnitude close to the mass shedding limit 
and can point in any direction.
\end{abstract}

\vspace{2pc}


\section{Introduction}
\label{sec:introduction}

\ac{BNS}~systems are very common in our Universe. Estimates suggest
their merger rate falls between $250$ to $2810$
Gpc$^{-3}$yr$^{-1}$~\cite{Abbott:2020uma}.
The coalescence of these \ac{BNS} systems is a source of
myriad phenomena such as, among others,
ejecta~\cite{Chaurasia:2018zhg, Chaurasia:2020ntk},
accretion disk~\cite{Fernandez:2018kax},
jets~\cite{Sun:2022vri},
r-process nucleosynthesis~\cite{Siegel:2022nature}, and
kilonova~\cite{Fernandez:2015use}.
These phenomena are treasure trove of information revealing
aspects of physics at large scales, for instance, gravity in strong regimes and
physical constants of the Universe, 
as well as aspects of physics regarding small scales,
like, \ac{EOS} in supranuclear dense matter and formation of heavy
elements in the periodic table.

As such, current detectors such as LIGO~\cite{LIGOScientific:2014pky},
VIRGO~\cite{VIRGO:2014yos}, and KAGRA~\cite{Kagra:10.1093} and the next
generation ones like
Cosmic Explorer~\cite{Reitze:2019iox},
the DECi-hertz Interferometer Gravitational-wave Observatory
(DECIGO)~\cite{Kawamura:2020pcg},
Einstein Telescope~\cite{Punturo:2010zz},
LIGO Voyager~\cite{Adhikari:2019zpy},
the Laser Interferometer Space Antenna (LISA)~\cite{LISA:2017pwj},
NEMO~\cite{Ackley:2020atn}, and
TianQin~\cite{TianQin:2015yph}
are designed to look eagerly into sky and observe the physical signals
emitted from the coalescence of compact binaries.

To unlock the wealth of information encoded in gravitational waves and 
their electromagnetic counterparts, accurate theoretical models are crucial. 
These models are essential for understanding events like 
GW170817~\cite{TheLIGOScientific:2017qsa}, 
the short gamma-ray burst GRB170817A~\cite{LIGOScientific:2017zic}, 
and the kilonova transient AT2017gfo~\cite{Coulter:2017wya}. 
However, finding analytical solution of the governing \acp{PDE},
when \ac{BNS} systems are coalescing, is not feasible as the \acp{PDE} are in a
highly non-linear regime where no approximation is
applicable~\cite{Foucart:2022iwu}.
In light of this, \ac{NR} community have put significant efforts to
solve these equations numerically and hence make sense of the observations.

Simulation of compact binary system in \ac{NR} often involves two steps. The
first step is to find the solution of Einstein-Euler \acp{PDE} on a
hypersurface of the spacetime manifold, namely, constructing constraint
satisfying and self-consistent \ac{ID} 
that present the binary system of interest at some
time. \ac{ID} codes such as,
{\tt COCAL}~\cite{Uryu:2011ky, Tsokaros:2015fea,Boukas:2023ckb},
\elliptica~\cite{Rashti:2021ihv},
{\tt FUKA}~\cite{Papenfort:2021hod},
{\tt LORENE}~\cite{Lorene, Grandclement:2006ht,
Taniguchi:2006yt,Taniguchi:2007xm, Taniguchi:2007aq},
{\tt NRPyElliptic}~\cite{Assumpcao:2021fhq},
{\tt SGRID}~\cite{Tichy:2009yr, Tichy:2012rp, Tichy:2019ouu},
{\tt SpECTRE}'s elliptic solver~\cite{Fischer:2021voj,Fischer:2021qbh},
{\tt Spells}~\cite{Pfeiffer:2002wt, Foucart:2008qt, Tacik:2015tja,
Tacik:2016zal},
{\tt TwoPunctures}~\cite{Ansorg:2004ds, Ansorg:2005bp},
are developed and utilized for this step.

The next step is to use the \ac{ID} and simulate the system's evolution over
time; hence we can find the solution over the spacetime of interest.
Dynamical evolution codes, for instance,
{\tt AthenaK}~\cite{Zhu:2024utz,Fields:2024pob},
{\tt BAM}~\cite{Bruegmann:2006ulg,Thierfelder:2011yi,Dietrich:2015iva},
{\tt BAMPS}~\cite{Bugner:2015gqa,Hilditch:2015aba,Renkhoff:2023nfw},
{\tt Dendro-GR}~\cite{Fernando:2018mov},
{\tt Einstein Toolkit}~\cite{Loeffler2012,EinsteinToolkit:2020_11},
{\tt ExaHyPE}~\cite{Koppel:2017kiz}
{\tt GR-Athena++}~\cite{Daszuta:2021ecf,Cook:2023bag,Rashti:2023wfe}
{\tt GRaM-X}~\cite{Shankar:2022ful,CarpetX_web},
{\tt GRChombo}~\cite{Clough:2015sqa,Andrade:2021rbd},
{\tt Nmesh}~\cite{Tichy:2022hpa},
{\tt NRPy+}~\cite{Ruchlin:2017com},
{\tt SACRA-MPI}~\cite{Kiuchi:2019kzt},
{\tt Simflowny}~\cite{Palenzuela:2018sly},
{\tt SpEC}~\cite{SpECWebsite,Boyle:2019kee},
{\tt SpECTRE}~\cite{Kidder:2016hev,Deppe:2021bhi}, and
{\tt SPHINCS\_BSSN}~\cite{Rosswog:2020kwm}, among others,
are made and employed for this step.

Previously, \elliptica~\cite{Rashti:2021ihv} was limited to 
the construction of \ac{ID} for black hole neutron star binary systems.
Additionally, the code was only supporting polytropic or piecewise polytropic \acp{EOS}.
In this work, we extend \elliptica's infrastructure to construction
\ac{ID} for \ac{BNS} systems as well as supporting tabulated \acp{EOS}, 
for instance, the
\compose~tables~\cite{Typel:2013rza,CompOSECoreTeam:2022ddl,Oertel:2016bki}.

To this aim, the Einstein-Euler equations are cast into
coupled non-linear elliptic \acp{PDE}~\cite{Tichy:2012rp,York:1998hy,Pfeiffer:2002iy},
and solved iteratively using Newton-Raphson method by the Schur domain
decomposition~\cite{Rashti:2021ihv}.
During the solve, the cubed spherical patches adapt to the surface of the
\acp{NS} thus separating matter and vacuum and preventing Gibbs phenomena in
the spectral method.
To achieve the target values of interest such as momenta of the system, 
mass and center of \acp{NS}, they are checked and adjusted during the solve.
The \acp{EOS} are approximated by either
(piecewise)~polytropic or tabulate ones. 

The remainder of paper is organized as follows.
In Sec.~(\ref{sec:formalism}) we explain the
mathematical background and formalism we use for \ac{ID} of \ac{BNS}
systems, in particular, Einstein-Euler equations and \ac{EOS} in \elliptica.
Sec.~(\ref{sec:numerical_method}) details the underling algorithms and
numerical techniques for construction of \ac{BNS}'s \ac{ID}.
We present the implementation of tabulated \ac{EOS} by spline interpolant means
and diagnostics in \elliptica.
Additionally, we explain the iterative procedure for finding
the physical and constraint satisfying \ac{ID} for \ac{BNS} systems.
In Sec.~(\ref{sec:results}),
we present various convergence tests and comparison against post
Newtonian answers to showcase the proof of concept for the new version the code.
In Sec.~(\ref{sec:summary}), we discuss the possible improvements and future work.

Throughout the article, we use geometric units where $G=1$ is the constant of 
gravity, $c=1$ is the speed of light, and solar mass $M_{\odot}=1$.
\added{Additionally, following~\cite{Read:2008iy}, we absorb the speed of light $c$
in the definition of pressure and total energy density.}

\section{Formalism}
\label{sec:formalism}
\subsection{Einstein-Euler equations}
\label{sec:einstein_euler_equations}

To derive the equations that govern the gravity and matter on a spatial-like hypersurface
$\Sigma_{t}$ of a 4-dimensional manifold $\mathcal{M}$ with a metric $g_{\mu\nu}$, 
we first write the line element of $\mathcal{M}$ as
\begin{equation}
  ds^2 = g_{\mu\nu}dx^\mu dx^\nu =
        -\alpha^2 dt^2 + \gamma_{ij}(dx^i+\beta^i dt)(dx^j+\beta^j dt).
\end{equation}
This specific form of the line element is particularly well-suited for the
$3+1$ formalism. It provides a clear view of the key variables~(fields):
$\alpha$, $\beta^i$, and $\gamma_{ij}$.
Here, $\alpha$ is the lapse gauge and indicates the way that
a sequence of spatial slices, $\Sigma_t$s, are combined to form the 
complete spacetime manifold $M$.
$\gamma_{ij}$ is the induced 3-metric on each spatial hypersurface and can be
written as $\gamma_{\mu\nu} = g_{\mu\nu} + n_\mu n_\nu$, where $n^\mu$
is the normal vector on $\Sigma_t$. $\beta^i$ is the shift vector.
The importance of this gauge, i.e., shift vector, is that it represents
the coordinate frame being used in each hypersurface.
In this work, we take the shift vector as follows
%
%
\begin{equation}
   \beta^i =
   B^{i} +
   \epsilon_{ijk} {\Omega}^{j}_{\mathrm{BNS}}(r^k-r^{k}_{\mathrm{CM}}) +
   \frac{v_r}{r_{\mathrm{BNS}}}(r^{i}-r^{i}_{\mathrm{CM}}).
   \label{eq:shift_def}
\end{equation}
Here, $\Omega^{i}_{\mathrm{BNS}}$ is the orbital angular velocity of the
\ac{BNS} system, $r^{i}_{\mathrm{CM}}$ is the position of the system's center of
mass, $r_{\mathrm{BNS}}$ is the coordinate distance between the \ac{NS} centers, 
$v_r$ is the radial velocity of the inspiraling coordinate system, and 
$\epsilon_{ijk}$ is the Levi-Civita symbol.
Eq.~(\ref{eq:shift_def}) proves
numerically convenient when applying \ac{BC} for $\beta^{i}$
at the edge of the computational grid where the position vector $\vec{r}$ 
has large values.

By definition the extrinsic curvature on $\Sigma_t$ is
$
  K_{\mu\nu} = -\frac{1}{2} \pounds_{n} \gamma_{\mu\nu},
$
in which $\pounds_n$ is the Lie derivative along the normal vector. We note
that by construction $K_{\mu\nu} n^\mu = 0$, therefore, we can use spatial~indices
to describe the extrinsic curvature.
Next, by utilizing conformal decomposition, we write
\begin{eqnarray}
  \gamma_{ij} = & \psi^4 \bar \gamma_{ij},
  \\
  K^{ij} = & A^{ij}+\frac{1}{3} K \gamma^{ij}.
\end{eqnarray}
Here $\psi$ is the conformal factor, $\bar \gamma_{ij}$
the conformal 3-metric, $A^{ij}$ the traceless part of $K^{ij}$.
Moreover, since we are using the \ac{XCTS}
formalism~\cite{York:1998hy,Pfeiffer:2002iy}

$A^{ij}$ is decomposed as
\begin{eqnarray}
   A^{ij} &= \psi^{-10} \bar A^{ij},
   \\
   \bar A^{ij} &= \frac{1}{2 \bar \alpha}
   \left((\bar L \beta)^{ij}-\bar \gamma^{ik}\bar
   \gamma^{jl}\bar u_{kl}\right),
\end{eqnarray}
where
\begin{eqnarray}
   \bar u_{ij} &= \frac{\partial \bar \gamma_{ij}}{\partial t},
   \\
   (\bar L \beta)^{ij} &= \bar D^i \beta^j + \bar D^j \beta^i
   - \frac{2}{3}\bar\gamma^{ij}\bar D_k \beta^k,
   \\
   \alpha &= \psi^6 \bar \alpha,
\end{eqnarray}
and $\bar D$ is the covariant derivative compatible
with $\bar \gamma_{ij}$.

To complete the formulation we need to incorporate the source terms.
We assume the fluid in \acp{NS} are governed by ideal fluid, hence the stress energy
tensor can be written as
\begin{eqnarray}
      T_{\mu\nu} & = &
           (\rho_0+\rho_0 \epsilon+P)u_\mu u_\nu + P g_{\mu\nu},
      \\ \nonumber
            & = & \rho_0 h u_\mu u_\nu + P g_{\mu\nu},
\end{eqnarray}
where, $\rho_0$ is the rest mass density,
$\epsilon$ the specific internal energy,
$P$ the pressure,
$h$ the specific enthalpy,
and $u^\mu$ the 4-velocity of the fluid.
Additionally, for $3+1$ decomposition purposes, we project the stress
energy tensor with respect to $\Sigma_{t}$ as follows
\begin{eqnarray}
  E &   = n_\mu n_\nu T^{\mu\nu}, \\
  S &   = \gamma^{ij}\gamma_{i\mu}\gamma_{j\nu}T^{\mu\nu}, \\
  j^i & = -\gamma^{i}_{\mu} n_{\nu}T^{\mu\nu},
\end{eqnarray}
where,
$E$ is the measured energy by the Eulerian observer whose 4-velocity is $n^\mu$.
$S$ is the trace of matter stress tensor,
and $j^i$ is the momentum flux.

Unique answer to a linear elliptic equation with a source is guaranteed by  
the maximum/minimum principle.
This principle becomes important during solve of constraint equations for
high mass \acp{NS}~(for further discussion see~\cite{Gourgoulhon:2007ue}).
To maintain the maximum/minimum principle, 
we rescale the stress energy projections as
\begin{eqnarray}
  E   & =  \psi^{-6} \bar E,
  \\
  S   & =  \psi^{-6}\bar S,
  \\
  j^i & =  \psi^{-6}\bar j^i.
  \label{eq:def_stress_energy_rescale}
\end{eqnarray}

Finally, following \ac{XCTS} formalism, 
we write Einstein's equations in quasi equilibrium condition
\begin{eqnarray}
   \bar D^{2}\psi-\frac{1}{8}\psi \bar R-
   \frac {1}{12}\psi^{5}K^{2}+
   \frac{1}{8}\psi^{-7}\bar A_{ij}\bar A^{ij}
   +2\pi \psi^{-1} \bar E &=& 0, \label{eq_psi}
   \\
   \bar D^{2}(\bar \alpha \psi^{7})-
   (\bar \alpha \psi^{7}) \left[ \frac {1}{8}\bar R+\frac {5}{12}
   \psi^{4} K^{2}+\frac {7}{8}\psi^{-8}\bar A_{ij}\bar A^{ij}\right]
   \nonumber \\
   + \psi^{5}(\partial_{t} K-\beta^k \partial_{k} K)
   -2\pi \bar\alpha \psi^{5}(\bar E+2\bar S) &=&0,\label{eq_lapse}
   \\
   2\bar \alpha \left[\bar D_{j}(\frac {1}{2\bar \alpha}
  (\bar L \beta)^{ij})-\bar D_{j}(\frac {1}{2\bar \alpha }\bar u^{ij})
   -\frac {2}{3}\psi^{6}\bar D^{i}K\right]
   -16\pi \bar \alpha \psi ^{4}\bar j^i &=&0,\label{eq_beta}
\end{eqnarray}
with the \acp{BC}
\begin{equation}
  \label{eq_bc_inf}
  \lim_{r\to\infty}\psi = 1, \ \ \
  \lim_{r\to\infty}B^i_0 = 0, \ \ \
  \lim_{r\to\infty}\alpha\psi = 1.
\end{equation}
Additionally, we pick the free data as
\begin{eqnarray}
  \bar \gamma_{ij} &= \delta_{ij},
  \label{eq_free_data_gamma}
  \\
  K & = 0,
  \label{eq_free_data_K}
  \\
  \bar u_{ij} & = 0.
\end{eqnarray}

For hydrodynamic equations, following~\cite{Tichy:2012rp},
we decompose the fluid into two parts:
the rotational part of the fluid and is represented by a cross
product, and the irrotational part of the fluid that is represented by a
velocity potential~(see, e.g.,\cite{Rezzolla_rh_book}).
In particular, the rotational part, which represents the \ac{NS} spin, reads
\begin{equation}
   \label{eq_ns_spin_vector}
   w^i=\epsilon_{ijk}
   \Omega^j_{\mathrm{NS}}(x^k-x^{k}_{c}),
\end{equation}
here, $\Omega^j_{\mathrm{NS}}$ is a free parameter to set the spin
level, see Sec.~(\ref{sec:spin}), and $x^{k}_{c}$ denotes the \ac{NS} coordinate.
The irrotational part is shown by the potential $\phi(x,y,z)$ and obeys 
the following equations.~\cite{Tichy:2019ouu}
\begin{eqnarray}
   & \frac{c\left(\rho_0\right)\alpha}{h} \psi^{-4} \bar\gamma^{ij}
   \partial_i \partial_j \phi
    - \frac{\rho_0\alpha}{h} \psi^{-4} \bar\gamma^{ij}
   \bar \Gamma^k_{ij}\partial_k \phi
   + 2\frac{\rho_0\alpha}{h} \psi^{-5}
   \bar \gamma^{ij}(\partial_i\psi) (\partial_j\phi)
  \nonumber \\
   &
   + \left(D_i \frac{\rho_0 \alpha}{h}\right) \left(D^i \phi\right)
    D_i \left[ \frac{\rho_0 \alpha}{h}  w^i
   -\rho_0 \alpha u^0 (\beta^i + \xi^i) \right] = 0,
   \label{eq_phi}
\end{eqnarray}
where, $\partial_i$ denotes the spatial partial derivative with respect to 
Cartesian coordinate $x^i$ and $\xi^i$ is the Killing vector.
Here, following the smoothing method in~\cite{Tichy:2019ouu}
\begin{equation}
c(\rho_0) = \rho_0 + \bar{\epsilon} \rho_{0c}
\left(\frac{\rho_{0c}-\rho_0}{\rho_{0c}}\right)^4 ,
\end{equation}
in which, $\rho_{0c}$ is the values of $\rho_0$ at the \ac{NS} center,
and $\bar{\epsilon}$ is a constant number -- for which we generally use $0.1$.
Since the fluid is enclosed inside the patches that cover the \ac{NS}, 
we further need to impose a \ac{BC} on the \ac{NS} surface as
\begin{equation}
  D^{i}\phi D_i\rho_0+w^{i} D_{i} \rho_0-
  h u^0(\beta^i+\xi^i) D_i\rho_0 = 0.
  \label{eq:bc_phi}
\end{equation}
We note that since Eqs.~(\ref{eq_phi},\ref{eq:bc_phi}) only include
derivatives of $\phi$, we can not uniquely determine $\phi$.
Hence, we further demand that the value of $\phi$ at the \ac{NS} center
is a constant number -- like 0.

\subsection{Equation of state}
\label{sec:eos}
In order to close the system of equations we need an \ac{EOS}.
\elliptica{} deploys specific enthalpy, $h$, to create a link between 
the macroscopic properties of the matter and the gravity.
As such, if we have a piecewise \ac{EOS} we write
\begin{eqnarray}
    \rho_0(h) &= K_i^{-n_i}\left(\frac{h-1-a_i}{n_i+1}\right)^{n_i},
   \nonumber\\
    P(h) &= K_i^{-n_i}\left(\frac{h-1-a_i}{n_i+1}\right)^{n_i+1},
   \nonumber\\
    \epsilon(h) &= \frac{a_i+n_i(h-1)}{n_i+1},
 \end{eqnarray}
where $n_i=\frac{1}{\Gamma_i-1}$ is the polytropic index and 
$K_i$'s are specific to the given \ac{EOS};
$a_i$'s ensure the continuity of \ac{EOS}~\cite{Read:2008iy}
and are set as
\begin{eqnarray}
   a_0 &= 0,
   \nonumber \\
   a_i &=
   a_{i-1}+
   \frac{K_{i-1}}{\Gamma_{i-1}-1}\rho_{i}^{\Gamma_{i-1}-1}
   -\frac{K_i}{\Gamma_i-1}\rho_i^{\Gamma_i-1}.
\end{eqnarray}
%

Similarly, for tabulated \acp{EOS}
we represent $\rho_0 \left(h\right)$,
$P \left(h\right)$, and $\epsilon \left(h\right)$ as spline interpolants given
discrete data points $\left( \rho_0, P, \epsilon, h \right)$. 
Although typically
only the quantities $\left( \rho_0, P, \epsilon \right)$ or 
$\left( \rho_0, P, e \right)$ are provided in tables, where $e$ is the total energy
density, we may convert between variables via the relations:
\begin{eqnarray}
    e &= \rho_0 \left( 1 + \epsilon \right),
    \nonumber \\
    h &= 1 + \epsilon + \frac{P}{\rho_0}.
\end{eqnarray}

Having written thermodynamic variables in terms of specific enthalpy,
we now write specific enthalpy in terms of the metric and fluid variables~\cite{Tichy:2012rp}
\begin{eqnarray}
   h &=& \sqrt{L^2 - (D_i \phi + w_i)(D^i \phi + w^i)},
   \nonumber \\
   L^2 &=& \frac{b + \sqrt{b^2 - 4\alpha^4 [(D_i \phi + w_i)
   w^i]^2}}{2\alpha^2},
   \nonumber \\
   b &=& [ (\xi^i+\beta^i)D_i \phi - C]^2 + 2\alpha^2 (D_i \phi + w_i) w^i.
   \label{eq_enthalpy}
\end{eqnarray}
%

\section{Numerical method}
\label{sec:numerical_method}

\subsection{Coordinate system}
\label{sec:coordinate_system}

The computation grid is tiled by cubed spherical coordinate 
systems~\cite{RONCHI199693}, except 
at the \ac{NS} centers, where we use simple Cartesian boxes to avoid
coordinate singularities at $r=0$~\cite{Rashti:2021ihv,Tichy_2019}. 
The relation between Cartesian coordinate, denoted by $x^i=(x,y,z)$, and 
cubed spherical coordinate, $X^i= (X,Y,Z)$, is 
\begin{eqnarray}
  X(x,y,z) = \frac{x}{z},\;
  Y(x,y,z) = \frac{y}{z},\;
  Z(x,y,z) = \frac{z-r_\mathrm{in}}{r_\mathrm{out}-r_\mathrm{in}},
  \label{eq_cubed_spherical_trans}
\end{eqnarray}
here, $X$ and $Y$ take value $\in[-1,1]$, and $Z\in[0,1]$.
Additionally, $r_{\mathrm{in}}$ and $r_{\mathrm{out}}$ are defined
\begin{eqnarray}
  r_{\mathrm{in}} &= \frac{\sigma_{\mathrm{in}}(X,Y)}{\sqrt{1+X^2+Y^2}},\;
  r_{\mathrm{out}} &= \frac{\sigma_{\mathrm{out}}(X,Y)}{\sqrt{1+X^2+Y^2}}.
  \label{eq_r_in-out}
\end{eqnarray}
where, the shape of the inner boundary of a patch along the radial direction 
is determined by $\sigma_{\mathrm{in}}(X,Y)$ and its outer boundary by 
$\sigma_{\mathrm{out}}(X,Y)$.
$\sigma(X,Y)$ is related to Cartesian coordinates 
by the equation $\sigma(X,Y) = \sqrt{x^2+y^2+z^2}$.
Finally, we note that while Eq.~(\ref{eq_cubed_spherical_trans}) is written for
patches along the z-axis, one can generalize this along any other axes.

As the distance from \acp{NS} increases,
we expect the fields fall as powers of $r^{-1}$.
To account for this behavior we use a new transformation for $Z$ coordinate, 
denoted by $\widetilde{Z}$, for the paches covering large radii of the grid.
The transformation reads
\begin{equation}
  \widetilde{Z} = \frac{\sigma_{\mathrm{out}}}
                 {\sigma_{\mathrm{out}}-\sigma_{\mathrm{in}}}
  \left(1-\frac{\sigma_{\mathrm{in}}}{r}\right),
\end{equation}
Here $r=\sqrt{x^2+y^2+z^2}$ and still $\widetilde Z\in[0,1]$.

Finally, \elliptica{} uses Chebyshev polynomials of the first kind 
for the basis of the spectral expansion and deploys the extrema of
Chebyshev polynomials for its collocation points~\cite{Rashti:2021ihv}.

\begin{figure*}[t]
	\centering
	\includegraphics[width=0.48\linewidth]{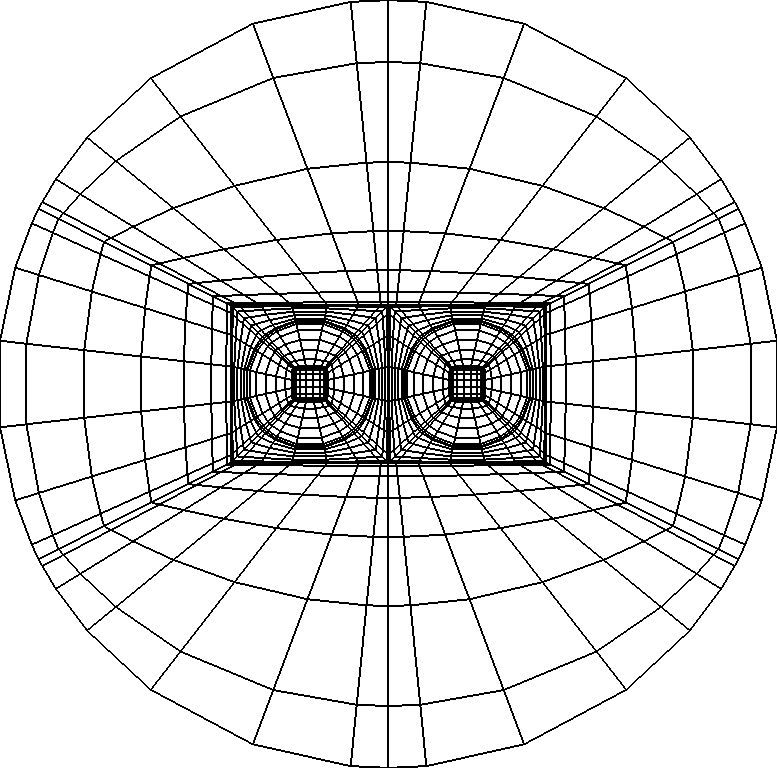}
	\includegraphics[width=0.48\linewidth]{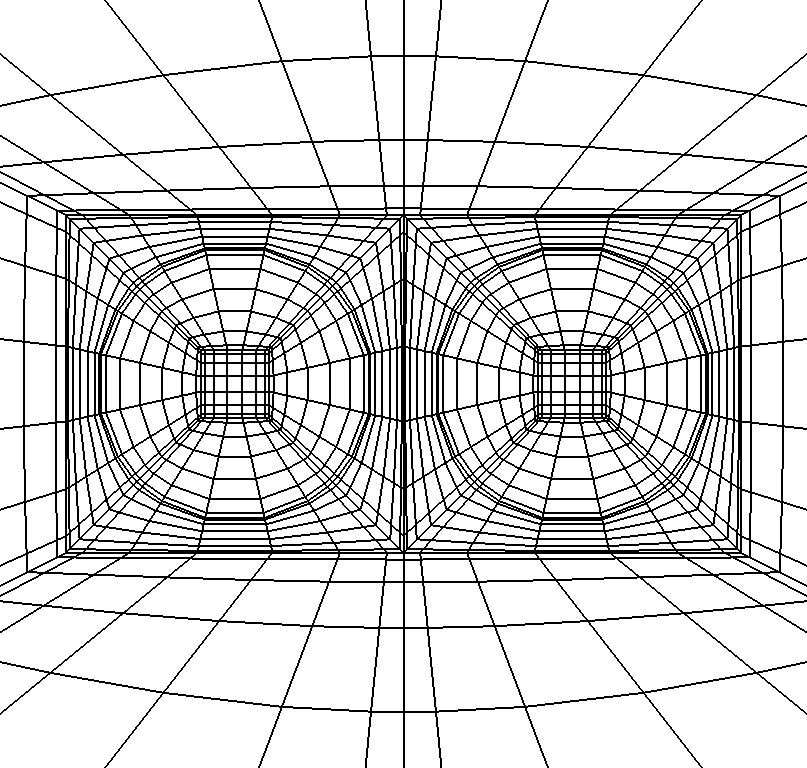}
	\caption{%
Grid patches for a \ac{BNS} system. 
Left: shown is the $x-y$ plane of the computational grid that is covered by
different patches.
\added{To fully represent the grid, the other boundary is set to a shorter length.
For a production run, the outer boundary is set generally at
a radius of $10^{5} \Msun$ from to the center of grid.}
Right: a zoomed-in view of the $x-y$ plane focusing on the \ac{NS} regions.
A Cartesian box is used around each \ac{NS} center to avoid coordinate singularities.
By using various $\sigma(X,Y)$ values, cubed spherical patches can adapt to different shapes, 
effectively capturing the shape of the NS surface. 
This allows for the treatment of matter and vacuum in separate patches.
}
	\label{fig:coord_system}
\end{figure*}

\fig{fig:coord_system} illustrates the patches covering the $x-y$ plane of
the computational grid used in this work.
\added{It is worth noting that the sizes 
used for the grid and \ac{NS} are reduced to illustrate the complete computational
grid. In practice, we set the outer boundary at a radius of $10^{5} \Msun$.}

\subsection{Elliptic solver}
\label{sec:elliptic_solver}

Given an elliptic \ac{PDE}, we linearize the equation to use the
Newton-Raphson method~(see~\cite{Rashti:2021ihv} for a complete description).
Thus, the original elliptic \ac{PDE} becomes a matrix equation, like $Ax=B$,
to be solved. Now the challenge is to solve this matrix equation
efficiently. To this aim, we use the Schur domain decomposition
method~\cite{Rashti:2021ihv,saad2003iterative}.
At the core of this method there is a divide and conquer strategy; in this
strategy, first the coupled equations are solved, and then the system decomposes 
into independent~(uncoupled) subsystems, amenable for parallel solving.
The Schur domain decomposition allows solving
the full system using shared memory multiprocessing in
which each CPU core is assigned to solve a matrix equations at each patch
independently.

The Schur domain decomposition method arranges 
the Jacobian matrix in Newton-Raphson method into two parts. 
The first part is associated with all uncoupled
equations and the second part comprises the coupled equations.
These couplings between equations occur due to the inter-\ac{BC}
at the interfaces of adjacent patches~\cite{Rashti:2021ihv}. 
As such, the system of matrix equations can be seen as 
two equations with two unknowns as follows
\begin{equation}
\label{eq_schur_2d}
 \left (
 \begin{array}{c c}
 B & E \\
 F & C
 \end{array}
 \right )
 \left (
 \begin{array}{c}
 \tilde v \\
 \tilde w \\
 \end{array}
 \right )
 =
 \left (
 \begin{array}{c}
 f \\
 g \\
 \end{array}
 \right ),
\end{equation}
where, the vector $\tilde v$ denotes all the unknowns that are uncoupled,
i.e., $\tilde v$ comprises all unknowns that stem from inner points of a patch
and hence uncoupled from the other patches.
The vector $\tilde w$ is the unknowns that are due to coupled equations,
i.e., stem from inter-\ac{BC} between adjacent patches.
By such an arrangement, if the unknown $\tilde w$ is solved, then
the equation of $\tilde v$ can be solved consequently.

Therefore, we first solve for $\tilde w$, and then we use it to solve for $\tilde v$.
Accordingly, in \elliptica, we solve \eq{eq_schur_2d} like:
\begin{eqnarray}
	\label{eq_syi_eq}
	(C - F E^\prime )\tilde w &= g - F f^\prime,\\
	\label{eq_xi_eq}
	\tilde v & = f^\prime -E^\prime \tilde w, 
\end{eqnarray}
where
\begin{eqnarray}
	E^\prime &= B^{-1} E, \nonumber \\
	f^\prime &= B^{-1} f.
	\label{eq_Ep_fp}
\end{eqnarray}
As we can see eq~(\ref{eq_syi_eq}) only involves the unknown $\tilde w$,
and it can be summarized as
\begin{equation}
	\label{eq_schur_y}
	S \tilde w = g^\prime ,
\end{equation}
where, $S$ is called Schur complement matrix.

After solving Eq.~(\ref{eq_schur_y}) for $\tilde w$, 
we solve for $\tilde v$ using Eq.~(\ref{eq_xi_eq}) -- hence we find
the solution of the whole system.

\begin{figure}[t]
\begin{algorithm}[H]
\caption{Schur complement domain decomposition method.}
\label{algo_schur}
\begin{algorithmic}[1]
	\State Solve $B E^\prime = E$ for $E^\prime$;
	\State Solve $B f^\prime = f$ for $f^\prime$;
	\State Compute $g^\prime = g-Ff^\prime$;
	\State Compute $S=(C-FE^\prime)$;
	\State Solve $S \tilde w = g^\prime$ for $\tilde w$;
	\State Compute $\tilde v=f^\prime -E^\prime \tilde w$;
\end{algorithmic}
\end{algorithm}
\end{figure}

A summary of the Schur complement domain method to solve elliptic equations is
shown in algorithm~\ref{algo_schur}. 
For a more in-depth discussion about implementation of Schur domain
decomposition and parallelization of matrix solver the reader may 
consult~\cite{Rashti:2021ihv}.
Finally, we use the publicly available and open-source 
\texttt{UMFPACK} direct solver~\cite{umfpack}.

\subsection{Diagnostics}
\label{sec:diagnostics}

In {\tt Elliptica} we compute the baryon mass of each \ac{NS}
using~\cite{gourgoulhon_thebook} 
\begin{equation}
        M_B = \int_{\mathrm{NS}} \rho_0 \alpha 
              \psi^6 \sqrt{\bar{\gamma}} d^3x,
        \label{eq_baryonic_mass_simplified}
\end{equation}
in which $\bar{\gamma}$ is the determinant of $\bar{\gamma}_{ij}$ and the
integration is taken over the volume of the \ac{NS}.

To measure the \ac{NS} spins, we can use the flat space coordinate rotational
Killing vector, following~\cite{Campanelli_2007}, on the surface of \ac{NS}:
\begin{eqnarray}
        \vec{\phi}_x & = -(z-z_c)\vec{\partial}_y+(y-y_c)\vec{\partial}_z,
        \nonumber
        \\
        \vec{\phi}_y & = +(z-z_c)\vec{\partial}_x+(x-x_c)\vec{\partial}_z,
        \nonumber
        \\
        \vec{\phi}_z & = -(y-y_c)\vec{\partial}_x+(x-x_c)\vec{\partial}_y,
\end{eqnarray}
in which $(x_c,y_c,z_c)$ is the coordinate center of the \ac{NS}
and $(\vec{\partial}_i)_{i \in \{x,y,z\}}$ are the basis vectors associated
with the Cartesian coordinates. 
Accordingly, \ac{NS} spin $S_i$, for the $i$ direction, is
measured on the surface of \ac{NS} by
\begin{equation}
        S_i = 
        \frac{1}{8\pi}\oint_{\mathrm{NS}} (\vec{\phi}_i)^j s^k K_{jk}dA,
        \label{eq_evaluate_ns_spin}
\end{equation}
where, $s^k$ is the normal vector on the \ac{NS} surface.
The dimensionless spin is defined
\begin{equation}
	\chi_i := \frac{S_i}{M^2_{\mathrm{NS}}}.
	\label{eq:chi_def}
\end{equation}
Additionally, following~\cite{Tichy_2019}, we can first compute 
the angular momentum $J_i$, the center $R^i_c$, and 
linear momentum $P_i$ of the \ac{NS}, and then calculate
$S_i$ as follows
\begin{equation}
        S_i = J_i-\epsilon_{ijk} R^j_c P_k.
	\label{eq_JRP}
\end{equation}
%

Since the chosen free data $\bar{\gamma}_{ij}$ in 
Eq.~(\ref{eq_free_data_gamma}) satisfies the
quasi-isotropic gauge condition~\cite{gourgoulhon_thebook} and 
$K$ in Eq.~(\ref{eq_free_data_K}) meets the
asymptotic maximal gauge condition~\cite{gourgoulhon_thebook}, 
the ADM linear momentums and angular momentums of the system are 
defined~\cite{gourgoulhon_thebook}
\begin{eqnarray}
        P^{\infty}_i  & = \frac{1}{8\pi} \lim_{S_t \to \infty} 
        \oint_{S_t} (K_{jk}-K \gamma_{jk})(\vec{\partial}_i)^j s^k dA,
        \label{eq_linear_momentum}
        \\
        J^{\infty}_i  & = \frac{1}{8\pi} \lim_{S_t \to \infty} 
        \oint_{S_t} (K_{jk}-K\gamma_{jk})(\vec{\phi}_i)^j s^k dA,
        \label{eq_angular_momentum}
\end{eqnarray}
here,
\begin{eqnarray}
        \vec{\phi}_x & = 
         -(z-z_{\mathrm{CM}})\vec{\partial}_y+
         (y-y_{\mathrm{CM}})\vec{\partial}_z,
        \nonumber
        \\
        \vec{\phi}_y & = 
        +(z-z_{\mathrm{CM}})\vec{\partial}_x+
        (x-x_{\mathrm{CM}})\vec{\partial}_z,
        \nonumber
        \\
        \vec{\phi}_z & = 
        -(y-y_{\mathrm{CM}})\vec{\partial}_x+
        (x-x_{\mathrm{CM}})\vec{\partial}_y.
\end{eqnarray}

Lastly, to calculate the total ADM mass of the system, we use~\cite{gourgoulhon_thebook}
\begin{equation}
\label{eq_adm_mass}
 M_{\mathrm{ADM}} =
	\int_{\Sigma_t} \left[ \psi^5 E 
        + \frac{1}{16\pi} \left( \bar A_{ij} \bar A^{ij} \psi^{-7}
        - {\bar R} \psi - \frac{2}{3} K^2 \psi^5 \right) \right]
        \sqrt{\bar \gamma} d^3 x.
\end{equation}

\subsection{Tabulated equations of state}
\label{sec:tab_eos}

We can import tabulated \acp{EOS} from the \compose~repository
\cite{Typel:2013rza,CompOSECoreTeam:2022ddl,Oertel:2016bki}. A general
\compose~table is first restricted to $T=0$. The baryon density $n_b$ is converted
to the rest mass density $\rho_0$ via the neutron mass $m_n$: $\rho_0 = m_n n_b$.

We numerically differentiate and interpolate the \ac{EOS} table
to generate a Hermite spline representation of the functions 
$\rho_0 \left(h\right)$, $P \left(h\right)$, and $\epsilon \left(h\right)$
(see Sec.~(\ref{sec:Appendix_Interpolation})). In practice,
the logarithms of these quantities are actually used to generate the
interpolants.

A number of slight
modifications may be made to the \ac{EOS} in preprocessing
in order to make it more amenable to interpolation. These modifications
are necessary for both the physical consistency of the \ac{EOS} and
to facilitate the convergence of the solution.

The first modification involves adjusting specific enthalpy at the \ac{NS} surface
for the \ac{EOS} tables in which the specific enthalpy falls below $1$.
To this aim, 
\added{at the minimum pressure, 
which we consider occurring at the beginning of the table,}
we scale all $\rho_0$ points by the multiplicative constant 
$\eta = h_\mathrm{min}$ corresponding to the lowest value of $h$
in the table. Accordingly, $h$ points are scaled by the value
$\eta^{-1}$, i.e:
\begin{eqnarray}
   \rho_0 \to \eta \rho_0
   \nonumber \\
   h = \frac{ e + P }{ \rho_0 } \to \frac{ e + P }{ \eta \rho_0 } = \eta^{-1} h.
   \label{eq_enthalpy_scaling}
\end{eqnarray}
This alteration can be carried through to dynamical evolution codes by
e.g. using the same \ac{EOS} table or by scaling the rest-mass
density\rmed{(or baryon number density)}\added{, at the minimum pressure,}
by the same factor \added{used in the preprocessing step.}

In addition, several other features of tabular \acp{EOS} may pose problems
for both interpolation and the use of specific enthalpy as the independent 
thermodynamic variable. Among these problems is the presence of a region 
near the surface of the \ac{NS} where both
$\frac{de}{dh}$ and $\frac{d\rho_0}{dh}$ diverge.
This region is common to
many \acp{EOS} with a `crust', including both tabular
\acp{EOS} such as SFHo~\cite{Steiner:2012rk} and piecewise
polytropics~\cite{OBoyle:2020qvf}. 
While this region is unavoidable if we wish to
accurately represent the equation of state via specific enthalpy, 
cf.~\cite{Lindblom:2010bb,Foucart:2019yzo}, it also limits
the accuracy of the solution within the \ac{NS}~(see \fig{fig:convergence}).

Another consideration is the spacing of the data points in the \compose~table,
which may be highly irregular especially for tables generated as a piecewise
combination of different models.
To re-grid the \ac{EOS}, we numerically differentiate the table points
using Fornberg's method, explained in Sec.~(\ref{sec:Appendix_Fornberg}),
and generate a low-degree Hermite interpolant,
Sec.~(\ref{sec:Appendix_Interpolation}),
which is sampled to produce new data points on an evenly-spaced grid. 
This method  decreases oscillations
when the new data points are themselves interpolated.
Additionally, the interpolation of the \ac{EOS} truncates the jumps in
derivatives of the thermodynamic variables, i.e., $\frac{de}{dh}$ and
$\frac{d\rho_0}{dh}$, that as we mentioned before, in some tables are not
well-defined. This low-degree Hermite interpolant
corresponds to approximating the derivatives as finite at these points 
(since infinite values would be unphysical anyway).

\begin{figure}[t]
    \centering
      \includegraphics[width=0.8\textwidth]{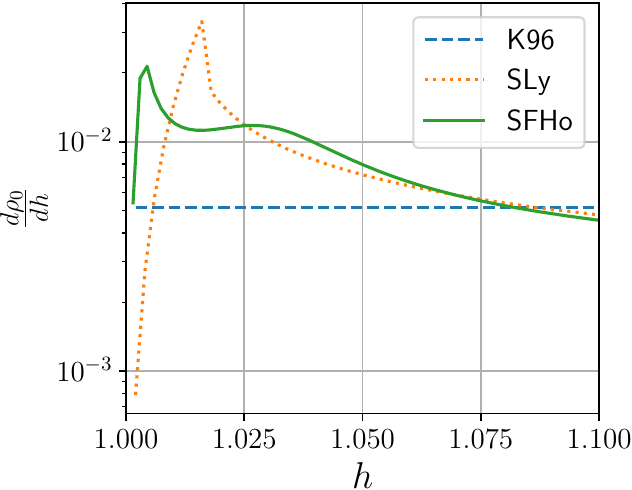}
    \caption{%
	Comparison of changes with respect to the specific enthalpy~($h$) 
	in \rmed{total energy density~($e$) and} the rest
	mass density~($\rho_0$) near the \ac{NS} surface for three
  \acp{EOS}: K96~(single polytrope), SLy~(piecewise polytrope), and SFHo~(table). 
  A lower specific enthalpy corresponds to a point closer to the \ac{NS} surface, 
	with $h=1$ defining the surface itself.
\added{$\rho_0$ is in geometric unit and $h$ is dimensionless.}
}
    \label{fig:eos_deriv_comparison}
\end{figure}

\fig{fig:eos_deriv_comparison} shows the resultant 
tabulated SFHo \ac{EOS} after re-griding using a low-degree Hermite interpolant
with comparison to \acp{EOS} of K96 and SLy.
The K96 \ac{EOS} is a single polytrope,
corresponding to $K = 96.7$ and $\Gamma = 2$~\cite{Baumgarte:2010ndz}.
SLy is an approximate piecewise polytrope of tabulated SLy4,
and therefore quantities such as \rmed{$de/dh$} $\frac{d\rho_0}{dh}$ 
are not continuous in certain regions corresponding
to transitions between piecewise segments~\cite{Haensel:2002cia}. SFHo,
here restricted to the zero-temperature
regime, shows similar discontinuities~\cite{Steiner:2012rk}.

Examining the functions relevant to the \ac{ID} generation, e.g.
$\rho_0$ and $e$,
we see that the single-polytrope model is considerably simpler and more
amenable to numerical solution
than the models with discontinuities. The difference is typically relevant
to the surface of the \ac{NS}
(where $h = 1$), as shown in the nearly vertical profile of 
$\frac{d\rho_0}{dh}$ \rmed{and $de/dh$} 
in \fig{fig:eos_deriv_comparison}. 
The sharp changes in $\frac{d\rho_0}{dh}$ \rmed{and $de/dh$}
with the piecewise polytropic and tabular \acp{EOS} are not perfectly 
captured by the grid, and thus they limit the convergence of the 
constraint violations (see Sec.~(\ref{sec:convergence_test})).

\added{
There are several potential ways to improve the convergence properties of the code 
in this area. For example, one could apply filtering to the 
spectral coefficients~\cite{jp_boyd_book} of $\rho_0$ or implement a routine to identify 
the location of discontinuities, thereby creating patches where the discontinuity 
occurs at the boundary between those patches.
Additionally, using spectral \ac{EOS}~\cite{Lindblom:2010bb, Foucart:2019yzo,Lindblom:2022mkr},
and imposing continuity at the piecewise polytrope \ac{EOS}~\cite{OBoyle:2020qvf}
have been shown to enhance the continuity of the solutions.
The implementation and comparison of these methods are left for future work.
}

\subsection{Initial data construction}
\label{sec:id_construction}

Construction of \ac{ID} often necessitates an iterative approach. This
process involves progressively refining the \ac{ID} 
until a suitable solution is achieved.
However, rapid or uncontrolled updates of the fields from one step to the next, 
or lack of adjustments of \ac{NS} masses or their centers can lead to a divergent 
answer and code crashes.
Another challenge is to find the \ac{NS} surface after each update,
as the true surface is unknown a-priori, 
and then creating a new computational grid with
patch's surfaces adapted to the new \ac{NS} surface.

We deploy the following iterative
procedure to construct \ac{ID} of a \ac{BNS} system.
In particular, we start from a coarse resolution grid and progressively
refine the answer, while controlling diagnostics, 
until the Hamiltonian and momentum constraints,
Eqs.~(\ref{eq_ham_constraint}) and (\ref{eq_mom_constraint}), reach a plateau.
Then, we increase the resolution and repeat this iteration until we achieve
the level of desired accuracy.


\textit{Step 0:}
We superimpose two \ac{TOV} star solutions as the initial guess
of the fields $\{\psi,\alpha\psi,B^i\}$. 
To initialize the $\phi$ fields for each \ac{NS} we use the approximation
$\phi = - \Omega^{z}_{\mathrm{BNS}}(y_{\mathrm{NS}}-y_{\mathrm{CM}})x$. 

\step{1}:
We solve the elliptic equations iteratively in this specific order:
first Eq.~(\ref{eq_phi}) for the matter field, and then
Eqs.~(\ref{eq_psi}), (\ref{eq_lapse}), and (\ref{eq_beta}) for the metric fields. 
This approach has been found to enhance the solution's stability, 
particularly when solving the matter field first. 
The order of solving the metric fields themselves, however, appears to be less critical.
Moreover, during each iteration step, we focus on solving a single elliptic equation 
while keeping the other fields fixed. In essence, the fixed fields act 
as source terms influencing the equation being solved.
Finally, within the Newton-Raphson iterative method, 
we perform only one update step per field, and then
incorporating the newly solved value into the source terms for the next equation.
In the following $\Xi$ denotes a field from the set
$\{\phi,\psi,\alpha\psi,B^i\}$.

\step{2}:
As mentioned earlier, 
iterative solvers are often sensitive to prompt changes; code crashes can
happen if a sudden update take place in the system. 
As such, we update the field solutions that are obtained from~\step{1} 
in a relaxed fashion. To this aim, we use
$\Xi = \lambda \Xi_{\mathrm{new}} + (1-\lambda) \Xi_{\mathrm{old}}$, in
which $\Xi_{\mathrm{old}}$ denotes the solution before entering
\step{1} and $\Xi_{\mathrm{new}}$ is the solution after exiting that step.
$\lambda$ denotes the weight of update. We
generally use $\lambda = 0.2$. This choice of $\lambda$ value is proven to work
for all experiments we have done.
   
\step{3}:
After updating all $\Xi$ fields, we see the baryon mass of each \ac{NS}
deviates, often by a few percents, from the target value.
Additionally, for spinning \acp{NS}, since at~\step{0} we used 
\ac{TOV} solution, and then we added the spin vector to the \ac{NS},
it is not surprising to observe the baryon mass is different from the
prescribed target value.
Moreover, the starting resolution is often coarse, and again 
the baryon mass may change as we go to a higher resolution.
If we do not account for these changes, 
the baryon mass deviates even further at later iterations and may lead to a
code crash.
To adjust the baryon mass, we note that $\rho_0 = \rho_0(h)$, and as shown in 
Eq.~(\ref{eq_enthalpy}) 
the specific enthalpy depends on a constant $C$, i.e., $h = h(C)$, which implies
$\rho_0 = \rho_0(C)$. 
Therefore by using a root finder, 
we find the value of $C$ in Eq.~(\ref{eq_baryonic_mass_simplified})
such that the baryon mass remains the same as the target value.

\step{4}:
Since at \step{0} we began with a rough approximation of the solution, 
the linear ADM momentums are initially not zero. Furthermore, for
asymmetric masses or spinning systems, we do not know in advance where is the exact position of
the system center of mass -- as one needs to consider the full general
relativity effect to find it. Therefore, we iteratively find the system
center of mass $\vec{r}_{\mathrm{CM}}$, by demanding the ADM momentum in each
direction to be zero.
The ADM momentum in $z$-direction proves consistently to be small, generally, 
$\frac{|P^{\infty}_z|}{M_{\mathrm{ADM}}} < 10^{-11}$, 
hence we keep $z_{\mathrm{CM}} = 0$, as the original value. 
However, often the initial value of ADM momenta in $x$ or $y$ direction is
$\approx 10^{-5} M_{\mathrm{ADM}}$.
Therefore, we adjust $x_{\mathrm{CM}}$ and $y_{\mathrm{CM}}$ as follows
\begin{eqnarray}
  x_{\mathrm{CM,new}} & =
  x_{\mathrm{CM,old}} + \lambda
  \frac{P^{\infty}_y}{\Omega^{z}_{\mathrm{BNS}} M_{\mathrm{ADM}}},
  \nonumber
  \\
   y_{\mathrm{CM,new}} & =
  y_{\mathrm{CM,old}} - \lambda
  \frac{P^{\infty}_x}{\Omega^{z}_{\mathrm{BNS}} M_{\mathrm{ADM}}}.
\end{eqnarray}
Here, $\lambda$ is generally chosen $0.2$.
It is worth noting that $K_{ij}$ is a function of $\beta^{i}$, while 
$\beta^{i}$ through Eq.~(\ref{eq:shift_def}) is a function of 
$r^{i}_{\mathrm{CM}}$. Therefore, by adjusting the $r^{i}_{\mathrm{CM}}$,
we influence $K_{ij}$ -- and hence all other fields as they are coupled to
another -- and we can derive $P^{\infty}_i$ to zero iteratively.

\step{5}:
We update the stress energy tensor, in particular, we
update the specific enthalpy for each \ac{NS} in a relaxed fashion as
$h = \lambda h_{\mathrm{new}}+(1-\lambda)h_{\mathrm{old}}$;
$\lambda$ is usually set to $0.5$. Then, we use the new value of specific enthalpy to
update $\rho_0(h)$, $\epsilon(h)$, and $P(h)$.

\step{6}:
Finding the orbital angular velocity, $\Omega^{z}_{\mathrm{BNS}}$, for a
quasi-circular orbit demands full solution of general relativity. Since we
start with a Newtonian approximation for this value we need to refine it.
Following~\cite{Tichy:2016vmv}, so called the force balance method,
we first compute $\partial_i \ln \Gamma$ at each \ac{NS} center, 
where $\Gamma$ computed as
\begin{equation}
  \Gamma =
  \frac{\alpha u^0 [1-(\beta^i+\xi^i+\frac{w^i}{hu^0})
    \frac{D_i \phi}{\alpha^2 h u^0}-
    \frac{w_i w^i}{(\alpha^2 h u^0)^2}]}
    {\sqrt{1-(\beta^i+\xi^i+\frac{w^i}{hu^0})(\beta_i+\xi_i
    +\frac{w_i}{h u^0})
    \frac{1}{\alpha^2}}}.
	\label{eq_fb_gamma}
\end{equation}
Then, we use a root finder to find $\Omega^{z}_{\mathrm{BNS}}$ such that
\begin{equation}
  \partial_i \ln[\alpha^2 -
  (\beta^i+\xi^i+\frac{w^i}{hu^0})
  (\beta_i+\xi_i+\frac{w_i}{h u^0})]
  + 2\partial_i \ln \Gamma = 0,
  \label{eq_force_balance}
\end{equation}
where, $\partial_i = \frac{\partial}{\partial x^i}$. 
Since in our setup the \ac{NS} centers sit on y-axis, 
we compute Eq.~(\ref{eq_force_balance}) along this axis.
We find $\Omega^{z}_{\mathrm{BNS}}$ for each \ac{NS} centers and update it
accordingly.

\step{7}:
At this point the specific enthalpy profile and hence \ac{NS} surface are changed.
We need to find the new location of \ac{NS} surface for adjusting
cubed spherical patches that are covering the \acp{NS}
-- so we can separate matter and vacuum into different patches.
Since some parts of the \ac{NS} may need to extend
to the patches that are currently covering vacuum, we 
extrapolate $h$ into these patches, so the root finder can find where $h=1$.
To this aim we use the following formula to extrapolate specific enthalpy to 
the vacuum
\begin{equation}
\label{eq_ns_extrap}
 f\left(r\right) = \left(a+\frac{b}{r}\right) \exp\left(-c_0\frac{r}{r_0}\right),
\end{equation}
here, $r$ denotes the coordinate distance from the \ac{NS} center, and $c_0$
is a constant~(typically $0.01$). 
We find the values of $a$, and $b$ by demanding $C^1$ continuity across the
\ac{NS} surface.
Additionally, we use the same \eq{eq_ns_extrap} to extrapolate $\phi$
field outside the \ac{NS}. This step is required when the \ac{NS} surface is
expanded, and we want to interpolate $\phi$ from the current grid to a new grid.

\step{8}:
We identify the center of each \ac{NS} by locating the coordinate 
where the specific enthalpy reaches its maximum value.
After multiple updates to the matter fields in previous steps, 
the \ac{NS} centers can exhibit slight drifts from their initial positions. 
These drifts can accumulate over time, potentially leading to code crashes.
To address this issue, we employ a corrective measure that adjusts the specific
enthalpy function as follows
\begin{equation}
  h_{\mathrm{new}}(\vec{r}) =
    h_{\mathrm{old}}(\vec{r}) -
    (\vec r - \vec r_0)\cdot\vec{\nabla} h_{\mathrm{old}}(\vec{r_0}),
\end{equation}
where, $r_0$ denotes the coordinate of \ac{NS} center.
This adjustment ensures that the maximum value of the specific enthalpy 
remains at the same location, 
effectively preventing the \ac{NS} centers from drifting significantly.

\step{9}:
We find the profile of $\sigma_{\mathrm{out}}(X,Y)$ that is necessary for
Eq.~(\ref{eq_r_in-out}) to have patch surfaces fitting the \ac{NS} surface.
To this end, for a given angular $\theta$ and $\phi$ in spherical coordinate, 
we use a root finder for $r$ to solve $h(r,\theta,\phi)-1 = 0$ and hence
finding the new \ac{NS} surface. Here, $r$ is the coordinate distance from
the \ac{NS} center.

\step{10}:
If the \ac{NS} surfaces are changed or 
if the resolution increases at the next iteration,
we create a new grid. To ensure a smooth transition, 
we use spectral interpolation techniques to transfer data 
from the previous grid onto the new one.

\step{11}:
We monitor the Hamiltonian and momentum constraints, 
Eqs.~(\ref{eq_ham_constraint}) and (\ref{eq_mom_constraint}).
We restart from \step{1} unless the constraints reach their truncation 
error and are level off. 
In this case, when the constraints reach a plateau,
we stop the iterative process if there is no higher
resolution demanded; otherwise we go to the next resolution and start from \step{1}.

\section{Results}
\label{sec:results}

\subsection{Convergence test}
\label{sec:convergence_test}

Since \elliptica{} is a pseudo-spectral code, the first expectation of the code
is spectral convergence feature. 
As such, we calculate the Hamiltonian and momentum constraints,
respectively, using
\begin{eqnarray}
   H & := R - K_{ij}K^{ij}+K^2 - 16 \pi E
   \label{eq_ham_constraint}
   \\
   M^i & := D_j (K^{ij} - \gamma^{ij} K ) - 8 \pi j^i
   \label{eq_mom_constraint}
\end{eqnarray}

For the convergence test, we generate \ac{ID} for two symmetric \ac{BNS} systems:
one with polytropic K96 \ac{EOS} and the other with tabulated SFHo \ac{EOS}.
In these systems, the \acp{NS} have baryon mass $1.4$ with no spin,
and their separation is $50 M_{\odot}$.

\begin{figure}[t]
    \centering
    \begin{subfigure}{0.45\textwidth}
      \includegraphics[width=\textwidth]{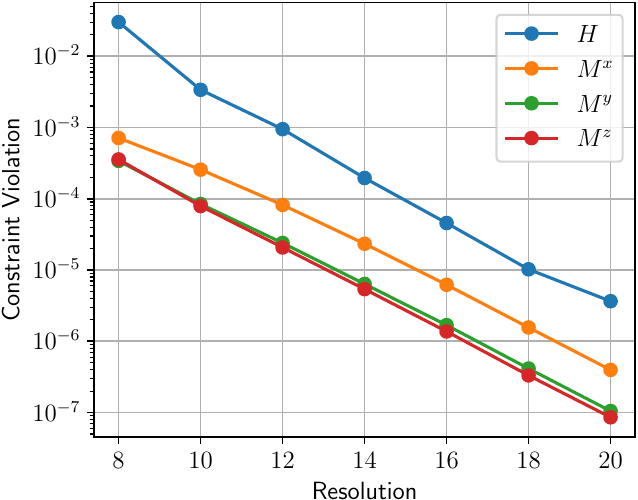}
      \caption{K96, \ac{NS} interior regions}
			\label{fig:k96_ns_int}
    \end{subfigure}
    \hfill
    \begin{subfigure}{0.45\textwidth}
      \includegraphics[width=\textwidth]{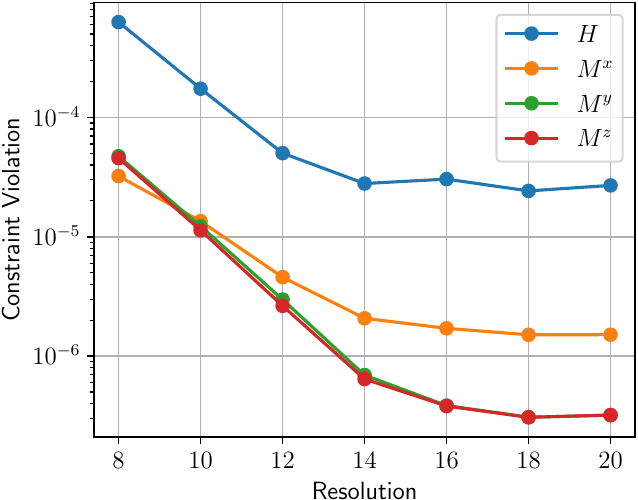}
      \caption{SFHo, \ac{NS} interior regions}
			\label{fig:sfho_ns_int}
    \end{subfigure}
    \hfill
    \begin{subfigure}{0.45\textwidth}
      \includegraphics[width=\textwidth]{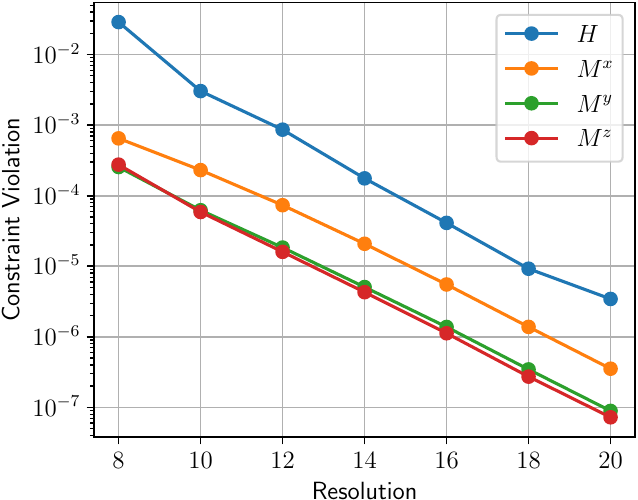}
      \caption{K96, \ac{NS} exterior regions}
		  \label{fig:k96_ns_ext}
    \end{subfigure}
    \hfill
    \begin{subfigure}{0.45\textwidth}
      \includegraphics[width=\textwidth]{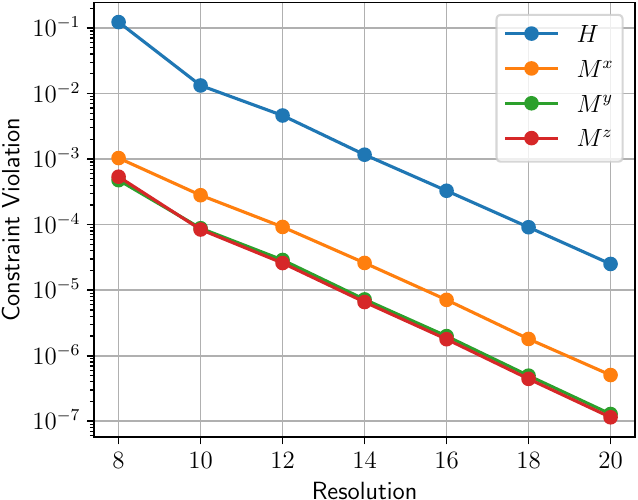}
      \caption{SFHo, \ac{NS} exterior regions}
			\label{fig:sfho_ns_ext}
    \end{subfigure}
    \begin{subfigure}{0.45\textwidth}
      \includegraphics[width=\textwidth]{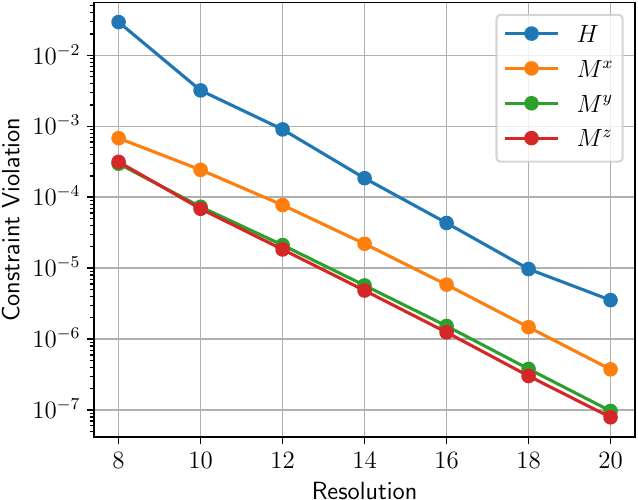}
      \caption{K96, all regions}
			\label{fig:k96_all}
    \end{subfigure}
    \hfill
    \begin{subfigure}{0.45\textwidth}
      \includegraphics[width=\textwidth]{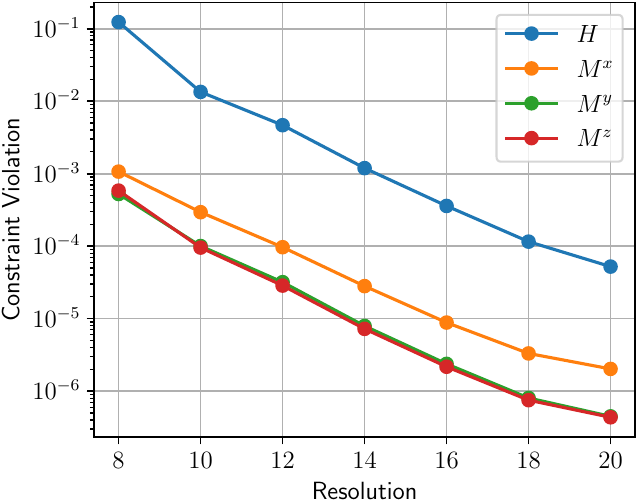}
      \caption{SFHo, all regions}
			\label{fig:sfho_all}
    \end{subfigure}
    \hfill

    \caption{The $L_2$-norm of the Hamiltonian and momentum constraint
    violations for the \ac{BNS} system with the K96 \ac{EOS}
    (single polytrope) and the SFHo \ac{EOS} (table). The constraint
    violations at each point are summed over all points in the specified
    regions. The regions not including a tabular \ac{EOS} show exponential
    convergence, while regions with such an \ac{EOS} reach a limit at
    high resolutions.}
    \label{fig:convergence}
\end{figure}

\fig{fig:convergence} shows the $L_2$-norm of the
Hamiltonian and momentum constraints as a function of grid resolution
(focusing on the final iteration at each resolution) for the two systems.
In figs.~(\ref{fig:k96_ns_int},\ref{fig:sfho_ns_int}),
the convergence test emphasizes on the \ac{NS}. 
We observe the constraint violations decrease exponentially for smooth
matter field, i.e., K96 \ac{EOS}. 
For the SFHo \ac{EOS}, initially for low resolutions
the constraints converge exponentially but for higher resolutions the 
constraints level off. 
This behavior is expected; when the grid resolution is coarse the
true discontinuities of \ac{EOS} variables are not seen by the spectral
expansion. For fine resolutions all features of \ac{EOS} variables
emerge and while spectral convergence try to resolve these features, but it is
not successful. Hence, we see constraints are soon level off and do not decrease
as we increase the resolution.
We note that, for the tabular \ac{EOS} (SFHo), the convergence is
limited due to the sharp features of $\frac{d \rho_0}{dh}$
\rmed{and $\frac{de}{dh}$} near the surface of the \ac{NS}
(see \fig{fig:eos_deriv_comparison}).
Indeed, the overall magnitude of the constraint violations is higher and
the rate of convergence is not exponential with
the tabular \ac{EOS} (in contrast to the simple polytrope).
Nonetheless, the profile of the convergence tests still suggest convergence
up to a limit at high resolutions.

Ensuring spectral convergence of the code for regions that fields are
smooth, we calculate $L_2$-norm of constraints at immediate neighboring
regions of \acp{NS}, where there is no matter fields.
Figs.~(\ref{fig:k96_ns_ext}, \ref{fig:sfho_ns_ext}) demonstrates they
convergence spectrally.

Finally, the overall converge of the constraints, calculated over all regions,
are shown in figs.~(\ref{fig:k96_all}, \ref{fig:sfho_all}); 
while both systems exhibit convergence as the resolution increases, the rate of
convergence for \ac{BNS} with SFHo \ac{EOS} is smaller than K96 \ac{EOS}.

\subsection{Post-Newtonian test}
\label{sec:pn}

We calculate the binding energy $E_b$ of a symmetric \added{non-spinning} 
\ac{BNS} system with \ac{NS} baryon mass $1.4$ and tabulated SLy4 \ac{EOS} for 
varied separations of \acp{NS}. Here
$E_b = M_{\mathrm{ADM}} - M_{\infty}$ and
$M_{\infty} = M^1_{\mathrm{TOV}} + M^2_{\mathrm{TOV}}$; 
we use \eq{eq_adm_mass} to calculate $M_{\mathrm{ADM}}$ and 
$M^{1/2}_{\mathrm{TOV}}$ are the corresponding gravitational mass of
\acp{NS} in isolation -- found by a TOV solver. 

To determine $\Omega^{z}_{\mathrm{BNS}}$ for a specific separation, 
we employ an iterative approach based on the force balance method 
described in \eq{eq_force_balance}. 
To validate the accuracy of the generated \ac{ID} against 
the expected analytical values for significant separations, 
particularly when the \ac{BNS} system exhibits quasi-circular motion, 
we calculate the binding energy $E_b$ corresponding to the 
given $\Omega^{z}_{\mathrm{BNS}}$ utilizing a post-Newtonian formula%
\added{, the 4 post-Newtonian order approximation},
outlined in \cite{Blanchet:2013haa}, \added{Sec.~7.4}.
\added{Specifically, to compute the post-Newtonian values of $E_b$, we need
the reduced mass $\mu = M^1_{\mathrm{TOV}} M^2_{\mathrm{TOV}} / {M_{\infty}}$,
the symmetric mass $\nu = \mu / M_{\infty}$, and the dimensionless orbital
velocity $ x^{3/2}  = \Omega^{z}_{\mathrm{BNS}} M_{\infty}$.
}

\begin{figure}[t]
 \includegraphics[width=0.8\textwidth]{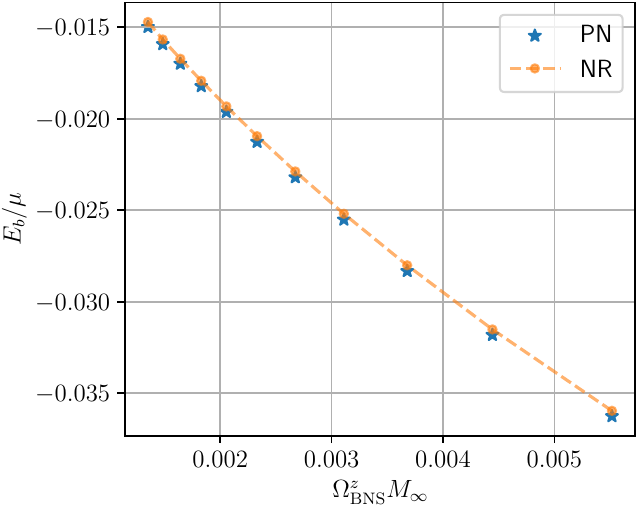}
 \caption{Analytic post-Newtonian~(PN) curve versus \ac{NR} curve.
	The values of $E_b$ for given $\Omega^{z}_{\mathrm{BNS}}$ are compared for
post-Newtonian method and \ac{NR} method. For binaries with a large
separation post-Newtonian and \ac{NR} values are matching. For \ac{BNS} of
a close separation there is a slight deviation from post-Newtonian prediction
as the system is not fully in quasi-circular status.
Here, $\mu = {M^1_{\mathrm{TOV}} M^2_{\mathrm{TOV}}}/{M_{\infty}}$ and the
binary has a symmetric baryon mass of $1.4$ for each \ac{NS}
and uses tabulated SLy4 \ac{EOS}. }
 \label{fig:pn}
\end{figure}

\fig{fig:pn} illustrates the binding energy of the \ac{BNS} 
system across different separations compared to the post-Newtonian data points. 
In cases of binaries with significant separations, meeting the quasi-circular 
motion criteria, we observe that the post-Newtonian and \ac{NR} values align. 
However, for smaller separations, the \ac{NR} values of $E_b$ are marginally higher.

\subsection{Spin}
\label{sec:spin}

We generate \ac{ID} for different 
$\Omega^{z}_{\mathrm{NS}}$, in Eq.~(\ref{eq_ns_spin_vector}),
values pertaining to one of the \ac{NS} in the \ac{BNS} system, 
where each \ac{NS} has a baryon mass of $1.4$, and they are separated by
$30 M_{\odot}$. In this scenario, we utilize the K96 \ac{EOS}.
The maximum dimensionless spin achievable is $\approx 0.56$, 
which corresponds to the mass shedding limit of a single \ac{NS} as 
discussed in~\cite{Ansorg:2003br}.
By increasing the $\Omega^z_{\mathrm{NS}}$ beyond $0.02$,
the spinning \ac{NS} becomes too oblate that the \ac{NS} surface finding
routine fails~(\step{9} in Sec.~(\ref{sec:id_construction})).

\begin{figure}[t]
 \includegraphics[width=0.8\textwidth]{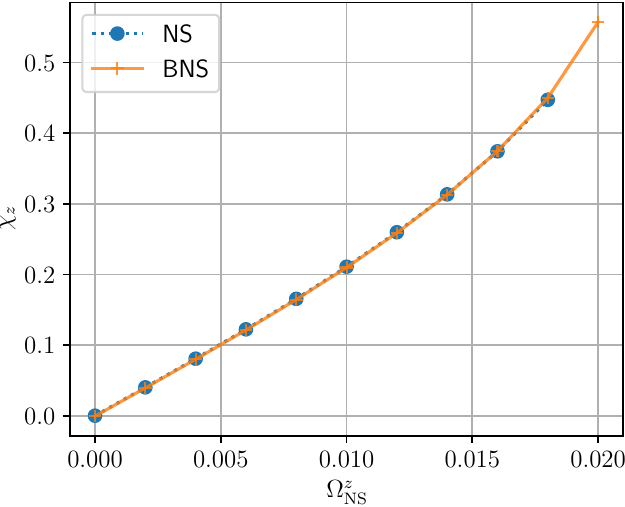}
 \caption{%
	The relation between $\Omega^z_{\mathrm{NS}}$ and $\chi_z$ is depicted 
	\added{by the solid line} for a 
	symmetric mass \ac{BNS} system with a baryonic mass of $1.4$ and a separation of 
  $30\Msun$. 
\added{The dashed line shows the corresponding values of $\chi_z$ for an
isolated spinning \ac{NS} for the same values of baryonic mass and \ac{EOS}.
We find an excellent agreement between spin values of the isolated \ac{NS} and 
the \ac{BNS} system.}
Initially, $\chi_z$ exhibits a linear growth pattern with 
  $\Omega^z_{\mathrm{NS}}$, but as $\Omega^z_{\mathrm{NS}}$ is raised 
  beyond a certain point, $\chi_z$ starts to increase non-linearly.
}

 \label{fig:spin}
\end{figure}

\added{
Additionally, to compare the spin values in a \ac{BNS} system with those of
an isolated spinning \ac{NS}, we create \ac{ID} for isolated \ac{NS} with the
same values of baryon mass, \ac{EOS}, and
$\Omega^{z}_{\mathrm{NS}}$ values~(except the largest value%
~\footnote{%
To create \ac{ID} for $\Omega^z_{\mathrm{NS}} = 0.02$, the \ac{NS} surface finding
routine fails, hence we can not report the corresponding spin value.}).
The resulting spin values are shown in \fig{fig:spin}, where we find
excellent agreement between spin value of the isolated \ac{NS} and 
the \ac{BNS} system. 
}

\fig{fig:spin} shows the relation between $\Omega^z_{\mathrm{NS}}$ 
in \eq{eq_ns_spin_vector} and $\chi_z$ in \eq{eq:chi_def}.
There is a linear \rmed{connection} \added{relation} between $\Omega^z_{\mathrm{NS}}$ and $\chi_z$ 
for low values of $\Omega^z_{\mathrm{NS}}$. 
However, as $\Omega^z_{\mathrm{NS}}$ increases, 
the relationship transitions into a nonlinear pattern, 
causing $\chi_z$ to rise more steeply.

Moreover, \fig{fig:spin} can be utilized as an approximate reference to 
determine the appropriate value of $\Omega^z_{\mathrm{NS}}$ 
corresponding to a desired spin level.

\subsection{\addedp{Outer boundary radius}}
\label{sec:bc_test}

We aim to study the effect of the outer boundary radius of the grid, 
where \eq{eq_bc_inf} is imposed, on the convergence behavior of the constraints 
and the binding energy of the system.
To this end, we construct four \ac{ID} sets for a symmetric non-spinning 
\ac{BNS} system with \ac{NS} baryon mass $1.4$ and the tabulated SLy4 \ac{EOS}
each featuring a different outer boundary radius. 
In particular, the chosen outer boundary radii are
$5 \times 10^{4} \Msun$, $1 \times 10^{5} \Msun$, 
$5 \times 10^{5} \Msun$, and $1 \times 10^{6} \Msun$.

We find that setting the outer boundary radius larger than $10^6 \Msun$, 
can lead to code failures due to issues with the patch finder routine in the
code. This routine is essential for establishing the 
inter-\ac{BC}~\cite{Rashti:2021ihv}
between adjacent patches, as it identifies neighboring patches for each patch.
Consequently, additional code support would be necessary if a larger outer boundary is required.
However, a grid with an outer boundary at $10^6 \Msun$ is already few orders of magnitude 
larger than the typical grids used in evolution simulations, 
such as those in~\cite{Cook:2023bag, Ujevic:2022qle, Bernuzzi:2024mfx}. 
Therefore, this size is sufficiently large for the \ac{ID} to be interpolated 
onto the grid of an evolution code.

\begin{figure}[t]
 \includegraphics[width=0.8\textwidth]{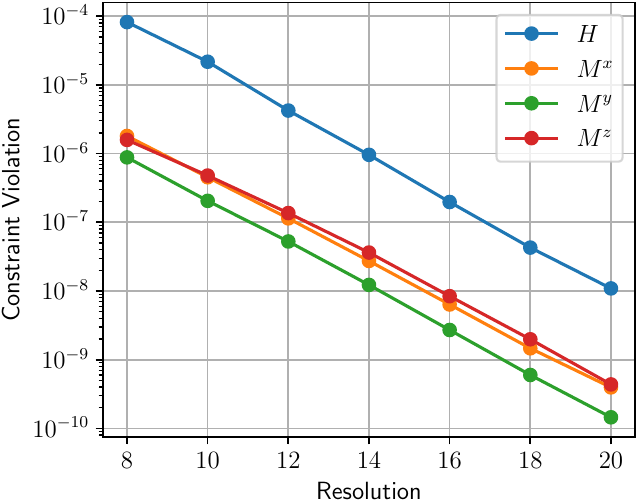}
 \caption{%
Shown is the $L_2$-norm of Hamiltonian~($H$) and momentum~($M^i$) constraint
violations for a symmetric \ac{BNS} system with the tabulated SLy4 \ac{EOS}
and the outer boundary radius of $5 \times 10^{5} \Msun$.
The constraints are calculated for an arbitrary patch that covers 
large-radius regions of the grid, where the flat spacetime \acp{BC},
\eq{eq_bc_inf}, are imposed on its outermost boundary.
Other patches within these large-radius regions display a similar convergence
pattern.
}
 \label{fig:bc_test_conv_test}
\end{figure}
\fig{fig:bc_test_conv_test} illustrates the convergence test of the Hamiltonian 
and momentum constraints focusing on one of the patches that 
cover the large-radius regions of the grid, i.e, reaching the outer boundary
of the grid.
We observe a clear spectral convergence for all constraints. 
This convergence behavior remains quantitatively unchanged across all other 
patches in these large-radius regions as well as across all other \ac{ID} sets with
different outer boundary radii.
Furthermore, the convergence behavior for
all regions, \ac{NS} exterior regions, and \ac{NS} interior regions 
are consistent with the findings for the SFHo \ac{EOS}, and discussed 
in Sec.~(\ref{sec:convergence_test}).

\begin{figure}[t]
 \includegraphics[width=0.8\textwidth]{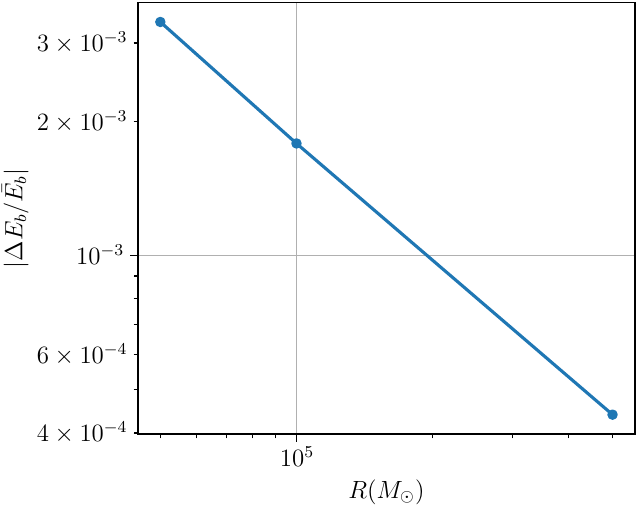}
 \caption{%
Relative difference in binding energy as a function of outer boundary radius.
We define $\Delta E_b = \bar E_b - E_b$, in which $\bar E_b$ represents the binding energy
of the \ac{BNS} system with the largest outer boundary radius, i.e., $ 10^6
\Msun$.
}
 \label{fig:bc_test_eb_test}
\end{figure}
Additionally, in \fig{fig:bc_test_eb_test}, 
we plot the absolute value of the relative difference in binding energy
as a function of the outer boundary radius. 
We define $\Delta E_b = \bar E_b - E_b$, in which $\bar E_b$ represents the binding energy
of the \ac{BNS} system with the largest outer boundary radius, namely, $ 10^6 \Msun$.
We observe that the relative difference in binding energy exhibits converging
behavior with respect to the $\bar E_b$ benchmark.
Presuming that the largest radius provides the most accurate binding energy~($\bar E_b$), 
we find the relative difference between the binding energies at smallest outer boundary radius
and the largest one is $\sim 0.3\%$. 
Furthermore, the relative difference in binding energies between 
$5 \times 10^{5} \Msun$ and $1 \times 10^{6} \Msun$ radii is
$\sim 0.04 \%$.

These results suggest that we can construct constraint satisfying and 
self-consistent \ac{ID} for different values of the outer boundary radius of
the grid.

\section{Summary}
\label{sec:summary}

In this work we have presented a significant upgrade to the \elliptica{}
infrastructure.
Previously limited to black hole-neutron star system and (piecewise)~polytropic 
\ac{EOS} for \acp{NS}, \elliptica{} can now construct \ac{ID} for 
spinning \ac{BNS} systems, 
incorporating realistic tabulated \acp{EOS} for the \ac{NS} matter.

To incorporate tabular \acp{EOS}, we have developed a number of techniques centered around
interpolation that allow convergence in the resulting solution
while remaining as true as possible to the underlying microphysics. While the
convergence of the solution is negatively affected by the complexity of the
\ac{EOS}, we maintain convergence up to a limit imposed by the surface features.

\added{
In this work, we selected tabulated SLy4 and SFHo \ac{EOS}s from the
\compose{} tables. Since these \acp{EOS} are generic among available cold \ac{NS} 
\ac{EOS} tables, our numerical methods remain still applicable to many such \acp{EOS}.
Additionally, our conclusions regarding convergence, for example, 
are relevant because the limiting factors on accuracy are common across these 
families of \acp{EOS}.
}

We have validated our code through convergence tests and 
comparisons with established analytical results, particularly in 
the post-Newtonian regime.
\addedp{
Additionally, by setting the outer boundary radius at different values, 
we study the convergence of the constraints and the binding energy of 
the system. We observe that the constraints exhibit spectral convergence 
in patches covering the large-radius regions of the grid. 
Furthermore, we find that choosing an outer boundary radius in the range of
$(10^5,10^6) \Msun$ results in an uncertainty bound of less than 
$\sim 0.2\%$ in the binding energy of the \ac{BNS} system.
}
These tests demonstrate the code's accuracy and robustness.

For smooth \ac{EOS}, such as polytropic models, 
we achieve spectral convergence, indicating an optimal error reduction rate 
with increasing resolution. However, for \ac{EOS} tables with discontinuities, 
spectral convergence is diminished. 
At high resolutions, we observe a convergence plateau inside \ac{NS}, 
though global convergence is still ensured. 
This behavior is expected for discontinuous functions.

\addedp{
While we have not studied our \ac{ID} in the context of an evolution simulation 
in this work, it is important to assess the quality of the \ac{ID} after its 
interpolation onto the grid of an evolution code. 
In particular, the propagation of constraint violations and the oscillation 
of the rest mass density at the center of each \ac{NS} are important
diagnostics to be studied.
Moreover, since different \ac{ID} codes employ different numerical parameters and algorithms
to solve \ac{ID} equations, it is important to examine the effects of these
differences on the resultant gravitational waves.
Specifically, from the point of view of the next generation detectors such
as Cosmic Explorer~\cite{Reitze:2019iox}, we need to study the mismatch between 
two gravitational wave signals emitted from the evolution of a \ac{BNS} system 
started from the same \ac{ID} but using two different \ac{ID} codes.
These topics are left for future work.
}

\added{
While some first steps have been taken to construct self-consistent 
\ac{ID} for magnetized \acp{NS} by 
solving the Einstein-Euler-Maxwell equations assuming ideal
magnetohydrodynamic~\cite{Gourgoulhon:2011gz,Tsokaros:2021pkh},
there are no \ac{ID} with self-consistent magnetic field for binary systems.
As such,
}
a particularly interesting area for future development lies 
in constructing self-consistent \ac{ID} for \ac{BNS} systems 
that contain magnetic fields.
This capability would be specially valuable for studying systems 
containing pulsars or magnetars, where these fields play a significant 
role in the dynamics.

\ack
AR and AN gratefully acknowledge David Radice for valuable discussions and 
insightful feedback on the manuscript.
AR acknowledges support from NASA under award No. 80NSSC21K1720.
The numerical simulations were performed on 
Roar Collab High Performance Computing Cluster at 
The Pennsylvania State University.

\elliptica{} is now public~\footnote{https://github.com/rashti-alireza/Elliptica} 
and open-source under \texttt{GPL-3.0} license.

\appendix

\section{Interpolation of the equation of state}
\label{sec:Appendix_Interpolation}

To generate a suitable interpolant of a tabular \ac{EOS},
we first approximate $\frac{\partial p}{\partial h}$,
$\frac{\partial \epsilon}{\partial h}$, and 
$\frac{\partial \rho_0}{\partial h}$
by finite-difference methods on the \compose{} data using Fornberg's
algorithm~\cite{fornberg_original}, explained in~\ref{sec:Appendix_Fornberg},
for unevenly-spaced grids.
Having the derivatives, we then generate a spline
interpolant composed of Hermite polynomials. 
The procedure for e.g. the pressure $p \left( h \right)$ is:

\step{0}:
We are given data points $(p_j,h_j)$ monotonically increasing in $h$.

\step{1}:
For each point $j$, we take the set of $N$ points
$X_j = \left\{ h_{j-\frac{N}{2}}, ..., h_j, ..., h_{j+\frac{N}{2}} \right\}$ 
centered around $h_j$. 
We shift the indices when needed for points near the boundaries $h_0$ and
$h_{j_{\mathrm{max}}}$.

\step{2}:
At each point, approximate 
$\left. \frac{ d p }{ d h } \right|_{h_j} \approx
 \sum_{\nu=j-\frac{N}{2}}^{j+\frac{N}{2}} \delta_{N,\nu}^{1} p_\nu$,
with the coefficients $\delta$ calculated by Fornberg's algorithm applied over the
set $X_j$ obtained in \step{1}.

\step{3}:
Having $p_j$ and its derivative $\frac{ d p(h_j) }{ d h }$,
we generate the interpolating Hermite polynomial (according to \cite{Burden:2011}) 
of desired order.

The same procedure is applied to obtain splines for $e \left( h \right)$,
$\rho_0 \left( h \right)$, and $\epsilon \left( h \right)$. Derivatives of these
functions are thereafter approximated by analytical derivatives of the spline interpolant.

\section{Application of Fornberg's finite difference method}
\label{sec:Appendix_Fornberg}

\step{2} of Sec.~(\ref{sec:Appendix_Interpolation}) entails finding derivatives
 such as $\frac{dp}{dh}$, which \ac{EOS} tables do not provide.
 We evaluate these derivatives numerically using finite difference
methods on the data points $\left( p_j, h_j \right)$. Since
the data points are typically unevenly spaced, we use Fornberg's finite difference
algorithm to generate finite difference coefficients that approximate
 $\frac{dp(h_j)}{dh}$ 
at each point~\cite{fornberg_original}. Specifically, the algorithm
calculates the weights $\delta_{N,\nu}^{m}$ such that
\begin{equation}
  \label{eqn:finite_difference}
  \left. \frac{ d^m p }{ {dh}^{m} } \right|_{h_j} \approx
  \sum_{\nu=j-\frac{N}{2}}^{j+\frac{N}{2}} \delta_{N,\nu}^{m} p_\nu,
\end{equation}
where $N$ is the number of points used (and determines the order of the 
finite difference approximation).

While Fornberg's algorithm can provide derivatives of arbitrarily high orders
(given enough data points), we are only interested in the first derivative. In
addition, we do not need the finite difference coefficients for all orders,
so we simplify the algorithm slightly. Then given a subset $X_j$ 
centered around $h_j$, from \step{1} of Sec.~(\ref{sec:Appendix_Interpolation}), 
we find the finite difference
weights using algorithm~\ref{algo_fornberg} (where $c_1$, $c_2$, and $c_3$ are introduced
just to simplify the notation, and $x_k$ is the $k$th element of $X_j$).
\begin{figure}[t]
\begin{algorithm}[H]
\caption{Simplification of Fornberg's finite difference algorithm, adapted
from~\cite{fornberg_original}.}
\label{algo_fornberg}
\begin{algorithmic}[1]
  \State $\delta^0_{0,0} = c_1 = 1$
	\For{$n=1,...,N$}
    \State $c_2 = 1$
    \For{$\nu = 0,...,n-1$}
      \State $c_3 = x_n - x_\nu$
      \State $c_2 = c_2 \cdot c_3$
      \If{$n=0,1$}
        \State $\delta^{n-1}_{n-1,\nu} = 0$
      \EndIf
      \State $\delta^{0}_{n,\nu} = \frac{ \left(x_n - h_j \right) }{c_3}
              \delta^{0}_{n-1,\nu}$
      \State $\delta^{1}_{n,\nu} = \frac{1}{c_3} \left( \left(x_n - h_j \right) 
              \delta^{1}_{n-1,\nu} - \delta^{0}_{n-1,\nu} \right)$
    \EndFor
    \State $\delta^{0}_{n,n} = \frac{c_1}{c_2} \left( h_j - x_{n-1} \right)
            \delta^{0}_{n-1,n-1}$
    \State $\delta^{1}_{n,n} = \frac{c_1}{c_2} \left( \delta^{0}_{n-1,n-1} 
            + \left( h_j - x_{n-1} \right) \delta^{1}_{n-1,n-1} \right)$
    \State $c_1 = c_2$
  \EndFor
\end{algorithmic}
\end{algorithm}
\end{figure}
%


\bigskip
\noindent
{\bf References\\}

\bibliographystyle{unsrt}

\end{document}